# X-Ray Propulsor: Physical Principle for an Electromagnetic Propellantless Propulsion System

**Alexandre A. Martins**

Institute for Plasmas and Nuclear Fusion & Instituto Superior Técnico,
Av. Rovisco Pais, 1049-001 Lisboa, Portugal
(aam@ist.utl.pt)

**Abstract.** In this work we are going to develop a physical model that explains how propulsion may be developed in a vacuum by the collision of electrons with an anode. Instead of using principles related to the conservation of only the mechanical momentum to achieve propulsion, like all the current propulsion systems do, the present system achieves propulsion by using principles related to the conservation of the canonical momentum. The complete physical model will be provided and comparison with preliminary experimental results will be performed. These results are important since they show that it is possible to achieve a radical different propulsion system with many advantages.

(Note to the readers: The propellantless physical principle exposed here is the basis of two submitted International patents (the earliest of which with a priority date from October 2010) detailing the experimental engineering and novel practical applications and configurations of this new and radical propulsion system where many variations and improvements are already being protected. These novel configurations will be published in due time.)

**1. Introduction**

Since the time of James Clerk Maxwell, scientists have been trying to unify all known forces (Waldrop, 2011) for the advancement of science and its benefits. The purpose of the present article is not to unify the known forces but to use one of them, the electromagnetic force, to achieve a form of propulsion that does not require the expulsion of mass from the system in one direction in order to obtain propulsion in the opposite direction. Lin *et al* (2011) explained how by using a time changing magnetic vector potential they were able to generate a synthetic electric field that generated a force on neutral atoms in a Bose-Einstein condensate, which behaved like charged particles in the presence of the generated induced electric field. We will demonstrate below how this induced electric field, from a time varying magnetic vector potential, (which satisfies both electromagnetic and mechanical momentum conservation laws) can be used to obtain a form of propellantless propulsion.

It started in the 1920's when Townsend Brown performed high voltage experiments in Coolidge vacuum tubes and capacitors at atmospheric pressures (Brown, 1928). He observed a directional force on all the setups that he attributed to a supposed electrogravitacional effect. Unfortunately he did not publish any paper on the subject or develop any theory. His latest efforts were all at atmospheric pressure (Brown, 1960; 1965) meaning that he didn't appreciate the value of the initial interesting results in vacuum tubes.

Later, his atmospheric setups prompted much curiosity which led several universities and institutions to try and replicate his results. It was found (Martins and Pinheiro, 2011a; 2011b) that the force responsible for propulsion at atmospheric pressure was a simple electrostatic interaction (mainly on the ground electrode) between the generated cloud of positive ions and the surrounding conductive armatures of the capacitor. In this case, the ions were generated by a corona discharge from a positively charged corona wire to a smooth ground electrode (figure 1). When compared to experiment (Chung and Li, 2007) our preliminary results in nitrogen (Martins and Pinheiro, 2011a) showed a small difference of only 12%.

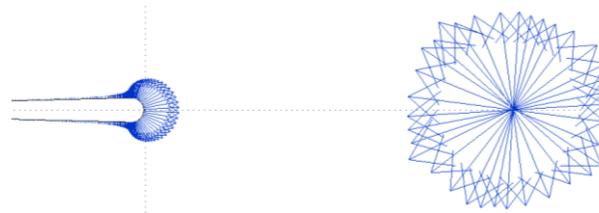

**Figure 1.** Electrostatic forces on the asymmetric capacitor used at atmospheric pressure.

NASA (Canning, Melcher, and Winet, 2004; Canning, Cole, Campbell, and Winet, 2004) tried to replicate Brown's results both in vacuum and at atmospheric pressures. They have provided measurements on the observation of an anomalous force in a high vacuum ($10^{-5}$Torr) associated with pulsed discharges or breakdowns which they recognized could not explain in conventional terms. Our own initial investigation (Martins and Pinheiro, 2011c) into this phenomenon led us to study the known vacuum arc thruster (VAT) which is known to create propulsive forces in a vacuum associated with pulsed arc discharges from the cathode to the anode (Tanberg, 1930; Robson and Engel, 1957; Engel and Robson, 1957).

Polk *et all* (2008) have reported on the highest possible theoretical VAT performance, according to the electrode material, as between 1.14 to 6.13 mN for pulsed discharges at 10 A. According to Lun (2008) the average directly measured thrust for a general VAT, without any external magnetic fields, generally increases linearly with the arc current (< 400 A) according roughly to 140 µN/A. Generally VAT operation induces erosion of the cathode electrode (Brown, 1990; Beilis, 2001; Amorim et al, 2001).

The force related by NASA as being associated with sparks in a vacuum (figure 2) was 0.014 N (Canning, Melcher, and Winet, 2004). NASA didn't provide any current measurements for the currents in the spark events, but they used a limiting resistor of 10 MΩ between the power supply and the capacitor in order to limit the current flowing through the circuit for safety purposes (Canning, Cole, Campbell, and Winet, 2004). The sparks were observed at 44 kV therefore the maximum current permissible was 4.4 mA.

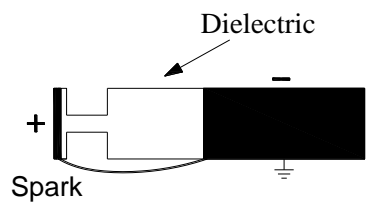

**Figure 2.** Spark from the cathode to the anode in the asymmetric capacitor used by NASA.

If the 14 mN force as its origin in VAT principles then the spark had to have a current of 100 A, inducing marked electrode erosion after repeated testing (Brown, 1990). Since erosion marks were not observed and the current (4.4 mA) was 22,727 times smaller than necessary, then a new model has to be provided to explain these results.

In order to try and understand how such a force could be generated a new model was made which explores the conditions in which the phenomenon takes place and if those conditions may give rise to an electromagnetic force not identified before and responsible for the observed force.

## 2. The Importance of the Right Kind of Spark Discharge

Although not apparent, there are many different types of spark discharges, all with different physical processes taking place (Boxman, Martin and Sanders, 1995). Chung and Li (2007) mention that whenever a spark discharge occurs in an asymmetric capacitor at atmospheric pressure it will lose all propulsive force. As we mentioned, the propulsive force that acts on these types of capacitors, in the atmosphere, has been proven to be of electrostatic origin (Martins and Pinheiro, 2011a; 2011b). In this case it is very important to have a space charge cloud between the corona wire and the smooth ground electrode, which induces a strong electrostatic attraction mainly on this last electrode. However, when a spark occurs the ionic cloud disappears and the electrostatic propulsion mechanism is interrupted.

The spark discharge of the VAT's in a vacuum is much different. In this case the currents are very high (tens of amperes to many kA) and the voltages are substantially lower (usually 30 V for a 200 A current and 3 kV just to initiate the arc). The electrode material contributes to the discharge itself where it is consumed in the process by thermal events. The current consists mainly of heavy elements and ions that generate propulsion by the conservation of mechanical momentum.

The high voltage discharge in a vacuum that occurs in experiments like NASA (44 kV) is again very different from the last two. In this case the voltage is much bigger than in the last case and surpasses the threshold ($10^8$ V/m) where the electrons can be removed from the cathode's surface and they are emitted in measurable quantities (Braithwaite, 2000). The current is small (4.4 mA) and constituted by the cathode ejected electrons only. In this case they suffer a big acceleration in the path between the electrodes because there are no particles for them to collide to. This process allows them to acquire a high energy which will be dissipated in a collision with the anode where very high induced electric fields and x-rays are generated, which are responsible for the force NASA observed. This process could not happen in the atmosphere due to the innumerous collisions the electrons make in their path generating ions and other metastable species. In this case the energy the electrons gain from the electric field is lost by this process and when they reach the anode they do not have the energy necessary to generate any electromagnetic propulsion force there. This is why sparks at atmospheric pressure inhibit the generation of any electromagnetic force on the electrodes. Therefore a high vacuum is necessary for this process to work.

## 3. Electromagnetic force calculation

As we mentioned, the spark discharge in a vacuum that we will consider occurs due to the removal of electrons from the cathode and their acceleration towards the anode. Since the currents are too small and the voltage too high, only electrons will compose this current. For the NASA experiment we have 44 kV applied over 6.35 cm (0.0635 m), creating an electric field of 692913.4 kV/m or 693 MV/m ($7 \times 10^8$ V/m), which surpasses the threshold ($10^8$ V/m) where the electrons can be removed from the cathode's surface with emission in measurable quantities (Braithwaite, 2000). The system we are considering is the asymmetric capacitor with a current flow (spark in this case) made only of electrons shown in figure 3.

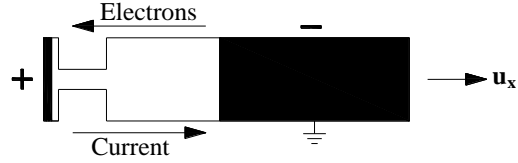

**Figure 3.** Schematic of the current generated by the electrons and axis system.

In our analysis we are going to use the known Euler-Lagrange equation for the total electromagnetic force on an electromagnetic system (Semon and Taylor, 1996):

$$\frac{d}{dt}(m\mathbf{v} + q\mathbf{A}) = -q\nabla V + q\nabla_{\mathbf{A}}(\mathbf{v}\cdot\mathbf{A}). \tag{1}$$

The canonical momentum $m\mathbf{v} + q\mathbf{A}$ is the quantity of momentum generally associated directly with the particles and fields (m**v** is the mechanical momentum and q**A** is the electromagnetic momentum). The two terms on the right depend on the gradient of the total interaction energy (electric and magnetic) between any charged particles and the surrounding particles and fields (the divergence operator $\nabla_{\mathbf{A}}$ acts only on the magnetic vector potential **A**), which in our case have a total zero resultant (the electrons emitted by the cathode slightly attract the anode and repel the cathode, but this electrostatic force is six orders of magnitude smaller than the main considered force). Since the two terms on the right are null or very small, the generalized or canonical momentum **p** will be conserved ($m\mathbf{v}+q\mathbf{A}=0$) and the electromagnetic force $\mathbf{F_{em}}$ (N) will become:

$$\mathbf{F_{em}} = d/dt(m\mathbf{v}) = -d/dt(q\mathbf{A}), \tag{2}$$

By derivation of this expression we obtain the force on the capacitor:

$$\mathbf{F_{em}} = \frac{d(m\mathbf{v})}{dt} = -q\frac{d\mathbf{A}}{dt} - \mathbf{A}\frac{dq}{dt}. \tag{3}$$

The second term on the right represents a force when the charge varies and is subject to a vector potential, but since we are considering an applied constant voltage (and that the power supply can provide the necessary current to maintain the voltage constant on the anode) it is not relevant for the force on the electrodes and will be discarded (however, if it is experimentally measured that with each spark the voltage on the anode varies considerably, then this term will be included). Considering that the charge $q$ is the charge $q_C$ on the capacitor plates, the total force on the electrodes will be:

$$\mathbf{F} = -q_C\frac{d\mathbf{A}}{dt} = q_C\mathbf{E}_{Induced}. \tag{4}$$

Where $\mathbf{E}_{induced}$ is the electric field produced by the time change of the vector potential **A** created by the accelerating or decelerating electrons. According to Jefimenko (1989) the external magnetic vector potential **A** of a current carrying wire of total length $L$ is given by:

$$\mathbf{A} = \frac{\mu_0 I}{2\pi}\ln\frac{L}{r}\mathbf{u_x}. \tag{5}$$

In this equation, $\mu_0$ is the magnetic permeability of the vacuum, $I$ is the electric current, $r$ is the radius of the current carrying wire and $\mathbf{u_x}$ is the unit vector in the positive direction as shown in figure 3. Equation (5) shows that the vector potential always has the direction of the current.

### 3.1. Charge on the NASA capacitor

The charge $q_C$ on a capacitor is given by:

$$q_C = CV, \tag{6}$$

where C is the capacitance and V is the potential difference or voltage between the two plates. The NASA capacitor is asymmetric, but since there is not a formula to calculate its capacitance directly we will use the known formula for the capacitance of a plane capacitor which is given by:

$$C = 8{,}84 \cdot 10^{-8} \cdot K \cdot \frac{S}{d}, \tag{7}$$

where $K$ is the relative dielectric constant of the dielectric, $S$ is the surface area (in cm$^2$) of the copper electrodes, $d$ is the distance between the plates (in cm), and $C$ is in $\mu F$. Simplifying the NASA capacitor we will consider that it is a plane capacitor with plate diameter of 3 cm, dielectric length of 6.5 cm and dielectric constant of 3 (which corresponds with data we could retrieve or guess from their reports). Therefore the NASA capacitor has approximately a capacitance of 2.88×10$^{-7}$ µF and when a voltage differential of 44 kV is applied between the plates it acquires a charge of 1.27×10$^{-8}$ C.

### 3.2. Force calculation

Using equations (4) and (5) we have:

$$\mathbf{F} = -q_C \frac{d\mathbf{A}}{dt} = -q_C \frac{d}{dt}\left[\frac{\mu_0 I}{2\pi}\ln\left(\frac{L}{r}\right)\right]\mathbf{u_x} = -q_C \frac{\mu_0}{2\pi}\ln\left(\frac{L}{r}\right)\frac{dI}{dt}\mathbf{u_x}. \tag{8}$$

Since $I = n_e q_e \mathrm{v}_e S,$ where $n_e$ is the number of charges per volume, $q_e$ is the charge of the involved electric charges, $\mathrm{v}_e$ is their velocity (in our case these charges will be the electrons) and $S$ is the effective area of the conducting path, the force will be:

$$\mathbf{F} = -q_C \frac{n_e q_e S \mu_0}{2\pi}\ln\left(\frac{L}{r}\right)\frac{d\mathrm{v}_e}{dt}\mathbf{u_x}. \tag{9}$$

Where $\mu_0$ is 1.2566 × 10$^{-6}$ (m kg C$^{-2}$), the length $L$ of the conducting path of the current is 0.0635 m, the radius $r$ of the conducting path is the electron radius (2.8178×10$^{-15}$ m), and the electron's charge is 1.6021×10$^{-19}$ C. We will take into account the relativistic increase of the electron's mass $m_{relativist}$:

$$m_{relativist} = \frac{m_e}{\sqrt{1 - \mathrm{v}_e^2/c^2}} = \gamma\, m_e. \tag{10}$$

Where $m_e$ is the rest electron's mass, $\mathrm{v}_e$ its velocity, $c$ is the velocity of light and $\gamma$ is the Lorentz gamma factor which determines the increase of the electron's mass with velocity. Considering that all the energy $q_e V$ is gained by the electrons and converted into the (relativistic) kinetic energy $(\gamma{-}1)m_e c^2$, then the electrons take on the velocity:

$$v_e = c\sqrt{1 - \frac{1}{\left(1 + \frac{q_e V}{m_e c^2}\right)^2}}. \qquad (11)$$

Taking into account this relativistic correction, the velocity reached by the electrons when arriving at the positive electrode in the considered case is $1.17 \times 10^8$ (ms$^{-1}$), which is 39% the velocity of light due to the low gamma factor of only 1.09. The mechanical momentum that the electrons transmit to the anode upon collision is extremely small when compared with the main electromagnetic force and will be given by (Kuhnen, Isoppo and Ouriques, 2007):

$$F = n_e m_e v_e = \frac{n_e m_e v_e}{\sqrt{1 - v_e^2/c^2}}. \qquad (12)$$

In the NASA experiment we know that the maximum current is 4.4 mA. Since the final velocity that the electrons acquire before collision is $1.17 \times 10^8$ (m/s), $n_e q_e S$ must be $3.76 \times 10^{-11}$ (C/m). According to our calculations the only relevant force is produced when the electrons are decelerating (because the deceleration is much bigger than the acceleration).

When the electrons decelerate they will generate the so called braking or Bremsstrahlung radiation. In the case of the NASA capacitor this will happen when they collide with the anode, decelerating from 39% the light velocity to zero velocity in a very short time. This process will also generate x-rays. The electron acquires an energy $E$ of 44 keV or $7.05 \times 10^{-15}$ J when subject to the 44 kV potential difference. Using the relation:

$$E = eV = h\nu, \qquad (13)$$

where $h$ is the plank constant ($6.6256 \times 10^{-34}$ Js) and $\nu$ is the frequency of the photon, we can determine the approximate value of the frequency of the generated x-rays as $1.06 \times 10^{19}$ s$^{-1}$. The inverse of this value is a measure of the time interval in which the electrons decelerate (when they collide with matter) and correspond to $9.40 \times 10^{-20}$ s. Using this last value and the velocity attained by the electrons at the anode (equation (11)) the deceleration will be $1.24 \times 10^{27}$ (ms$^{-2}$). Using this value in equation (9) we obtain for the force a value of $3.65 \times 10^3$ N. This would be the case if all the electron's energy were converted into x-rays, however only a very small fraction of its energy will actually be converted into x-rays, the rest being dissipated into heat and other electromagnetic radiation.

The efficiency $\eta$ of the x-ray production is defined as the ratio of the x-ray radiation to the full electron energy when it passes through the potential difference generated by the applied voltage V between the electrodes. It is determined by the type of deceleration fields and the depth to which the electrons penetrate the anode material, which is related to the atomic number $Z$. Therefore the x-ray production efficiency $\eta$ is given by (Hertrich, 2005):

$$\eta = k \cdot V \cdot Z. \qquad (14)$$

Where *k* is a constant of magnitude given experimentally as $1.1\times10^{-9}$. In the NASA capacitor we have applied 44 kV to a copper electrode of atomic number 29, which gives an efficiency of $1.4\times10^{-3}$ or 0.14% in these conditions. Due to these energy losses the final value for the calculated electromagnetic force will have to be calibrated by multiplying it with the x-ray production efficiency *η*, which gives a final value of 5.13 N for the electromagnetic force calculated before.

This would be the total force if all the electrons lost their energy in a full frontal collision, but since every electron is decelerated in different ways (have different trajectories relative to the target atoms), in bremsstrahlung, a continuous spectrum with a characteristic profile and energy cutoff (i.e., wavelength minimum, $\lambda_{min}$) is produced, known as white radiation (figure 4). Since the highest energy is given by:

$$E = eV = h\nu_{max} = hc/\lambda_{min}, \qquad (15)$$

then the minimum wavelength would be (Cullity, 1956):

$$\lambda_{min} = hc/eV. \qquad (16)$$

Therefore the minimum cutoff wavelength will be $0.28\times10^{-10}$ m or 0.28 Å. In order to obtain the total force from the colliding electrons, we would have to integrate the force for all the different electron energies (which correspond to different deceleration times). The maximum in the intensity of the white radiation spectrum generally occurs at a wavelength that is near $1.5\times\lambda_{min}$ or $0.423\times10^{-10}$ m. If all the electrons had this wavelength the force produced would be 3.42 N. At a wavelength of $10\times\lambda_{min}$ or $2.82\times10^{-10}$ m the total force would be 0.51 N.

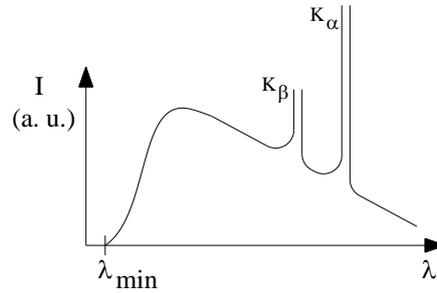

**Figure 4.** Typical x-ray spectra for copper with $K_\beta$ and $K_\alpha$ lines.

The calculations that have been performed until now are for constant operation of the electron emission and collision. The NASA capacitor operates with spark discharges of short duration and therefore the resultant force will be much smaller. However, we can try and compare the theoretical and experimental impulses generated.

As is known from capacitor discharge theory, the time to discharge to 99% is 5RC, where R is the limiting resistance and C is the capacitance. This gives the approximate value of 14.42 μs for the duration of the spark discharge. The impulse *J* delivered by a constant force *F* is $F\cdot\Delta t$, where *Δt* is the time interval during which the force is exerted. Since the measured force was 14 mN, then the experimental impulse must be around $2.02\times10^{-7}$ Ns. In table 1 it is shown the value of the theoretical impulses when all electrons have the respective wavelength shown and therefore represent only an upper limit for the developed impulse. We cannot compare directly these values with the "experimental" impulse value (we would have to perform a detailed integration for the whole spectrum), but the data indicates that it is possible that the order of magnitude for these impulses may be similar.

**Table 1.** Comparison of generated impulses.

|  | Impulse (Ns) |
|---|---|
| $\lambda_{min}$ | $7.40 \times 10^{-5}$ |
| $1.5 \times \lambda_{min}$ | $4.93 \times 10^{-5}$ |
| $10 \times \lambda_{min}$ | $7.35 \times 10^{-6}$ |
| NASA | $2.02 \times 10^{-7}$ |

We have considered the cathode in the NASA capacitor as being flat and parallel to the anode, and the dielectric between them as being continuous in our theoretical calculation for the anode capacitance. However, the cathode is a hollow cylinder and the dielectric has a big gap in the middle. These two factors will work to diminish the real capacitance/charge on the anode and therefore the force/impulse will be smaller than we have calculated like shown on table 1. On the other hand, the values in table 1 are for the case that all electrons have the respective wavelength shown, but (as mentioned) the force must be integrated over all energies, which will automatically diminish the calculated theoretical values for the propulsion force. Furthermore, the $K_\beta$ and $K_\alpha$ lines represent further losses in the generation of x-rays usable for propulsion which will have to be properly quantified. All these factors bring the theoretical and experimental values closer together.

Another important factor is that the presented theoretical calculations consider that the deceleration induced electric field acted on the whole of the charge present at the anode, but as we can see in figure 2, the spark will hit only at one extremity of the anode in the NASA case. This fact alone diminishes considerably the generated force making it harder to compare with the presented theory. In order to better compare experimental results with our theory it would be better if the electron current were constant and impacted the anode at the center. The two submitted International patents detail more efficient configurations like this (that may be inside a vacuum tube for ease of application) and also manages (among other things) to increase the force by increasing the capacitance/charge of the anode by several different means that will be detailed in the future.

### 4. Conclusion

We have demonstrated theoretically a new propellantless propulsion system that is converting the change of mechanical momentum $m\mathbf{v}$ (collision with the anode that induces a high magnitude $d\mathbf{v}/dt$, or deceleration $\mathbf{a}$) of the electrons, that is responsible for a massive change of the electrons electromagnetic momentum $q\mathbf{A}$ (high magnitude $d\mathbf{A}/dt$, which generates a very intense unidirectional electric field $\mathbf{E}_{ind}$), into a change of electromagnetic and mechanic momentum of the positive charge present at the anode. This generates a unidirectional electromagnetic propulsion force on the anode's charge that drags the whole setup that is mechanically connected. Using higher voltages and anode targets with high atomic numbers will be extremely advantageous because of the increased efficiency in x-ray production.

The value of the real capacitance, discharge current, duration of discharge current, generated force and a detailed experimental data on the resulting x-ray energy spectrum distribution must be acquired in order for more detailed calculations to be made in the future. With this in mind we are preparing detailed experimental measurements in the near future.

Although we have tried to present a model that includes all the expected physical processes that affect its performance, this is not a final model and will be adapted according to future needs. This model of the electromagnetic force generated by decelerating electrons (x-ray propulsor) on the positive charges present in the anode is able to correctly predict the expected magnitude of the impulse observed in the NASA experiment. The theory presented here also provides a very good understanding of the physical mechanism responsible for this new electromagnetic force and allows the future engineering of the effect for propulsion purposes.